\documentclass[twocolumn, tighten, times]{aastex631}

\shorttitle{JWST AbsFlux Program}
\shortauthors{Gordon et al.}

\begin{document}

\title{The James Webb Space Telescope Absolute Flux Calibration. I. Program Design and Calibrator Stars}

\author[0000-0001-5340-6774]{Karl D.\ Gordon}
\affiliation{Space Telescope Science Institute, 3700 San Martin
  Drive, Baltimore, MD, 21218, USA}
\affiliation{Sterrenkundig Observatorium, Universiteit Gent,
  Gent, Belgium}
  
\author[0000-0001-9806-0551]{Ralph Bohlin}
\affiliation{Space Telescope Science Institute, 3700 San Martin
  Drive, Baltimore, MD, 21218, USA}
  
\author[0000-0003-4520-1044]{G.~C.\ Sloan}
\affiliation{Space Telescope Science Institute, 3700 San Martin
  Drive, Baltimore, MD, 21218, USA}
  \affiliation{Department of Physics and Astronomy, University
  of North Carolina, Chapel Hill, NC 27599-3255, USA}

\author[0000-0003-2303-6519]{George Rieke}
\affiliation{Steward Observatory, University of Arizona, Tucson,
  AZ 85721, USA}

\author[0000-0002-3824-8832]{Kevin Volk}
\affiliation{Space Telescope Science Institute, 3700 San Martin
  Drive, Baltimore, MD, 21218, USA}
  
\author[0000-0003-4850-9589]{Martha Boyer}
\affiliation{Space Telescope Science Institute, 3700 San Martin
  Drive, Baltimore, MD, 21218, USA}

\author[0000-0002-5943-1222]{James Muzerolle}
\affiliation{Space Telescope Science Institute, 3700 San Martin
  Drive, Baltimore, MD, 21218, USA}

\author[0000-0001-8291-6490]{Everett Schlawin}
\affiliation{Steward Observatory, University of Arizona, Tucson,
  AZ 85721, USA}

\author[0000-0003-2823-360X]{Susana E.\ Deustua}
\altaffiliation{formerly at Space Telescope Science Institute}
\affiliation{Sensor Science Division, National Institute of Standards and Technology, 100 Bureau Dr., Gaithersburg, MD 20899-8444, USA }

\author[0000-0003-4653-6161]{Dean C.\ Hines}
\affiliation{Space Telescope Science Institute, 3700 San Martin
  Drive, Baltimore, MD, 21218, USA}

\author[0000-0002-2626-7155]{Kathleen E.\ Kraemer}
\affiliation{Institute for Scientific Research, Boston College, 140 Commonwealth Avenue, Chestnut Hill, MA 02467, USA}

\author[0000-0001-7106-4683]{Susan E.\ Mullally}
\affiliation{Space Telescope Science Institute, 3700 San Martin
  Drive, Baltimore, MD, 21218, USA}

\author[0000-0002-3532-5580]{Kate Y.\ L.\ Su}
\affiliation{Steward Observatory, University of Arizona, Tucson,
  AZ 85721, USA}


\begin{abstract}
It is critical for James Webb Space Telescope (JWST) science that instrumental units are converted to physical units.
We detail the design of the JWST absolute flux calibration program that has the core goal of ensuring a robust flux calibration internal to and between all the science instruments for both point and extended source science.
This program will observe a sample of calibration stars that have been extensively vetted based mainly on Hubble Space Telescope, Spitzer Space Telescope, and Transiting Exoplanet Survey Satellite observations.
The program uses multiple stars of three different, well understood types (hot stars, A dwarfs, and solar analogs) to allow for the statistical (within a type) and systematic (between types) uncertainties to be quantified.
The program explicitly includes observations to calibrate every instrument mode, further vet the set of calibration stars, measure the instrumental repeatability, measure the relative calibration between subarrays and full frame, and check the relative calibration between faint and bright stars.
For photometry, we have set up our calibration to directly support both the convention based on the band average flux density and the convention based on the flux density at a fixed wavelength.
\end{abstract}


\section{Introduction} \label{sec:intro}

The goal of absolute flux calibration is to convert astronomical measurements in instrumental units to physical units.
Answering most science questions requires measurements in flux densities or surface brightnesses.
While many science programs require 5--10\% flux calibration accuracy, higher accuracy significantly enhances supernovae, dark energy, stellar populations, and stellar structure investigations \citep{Kent09}.
A common method for such calibration is to measure the flux density of a star in instrumental units and compare it to its predicted flux density, which is often based on a model of a dust extinguished stellar atmosphere, with their parameters determined by fitting previous calibrated measurements \citep[e.g.,][]{Cohen99, Rieke08, Engelke10, Bohlin14}. 
While stars are the sources most often used for absolute flux calibration, other sources have been used including laboratory calibrated blackbodies \citep{Fixsen94, Price04} and asteroids \citep{Stansberry07, Muller14}.

The ideal flux calibration source is one that has been directly calibrated against laboratory standards at all wavelengths of interest.
A few sources have been calibrated this way, but only at specific wavelengths.
For example, the star Vega has been calibrated at the monochromatic wavelength of 5556~\AA\ through observations using ground-based telescopes and laboratory calibrated sources \citep[e.g.,][]{Megessier95}.
In the mid-infrared (MIR), the space-based Mid-Course Space Experiment (MSX) observed a number of bright stars and small calibrated spheres \citep{Price04}.
In particular, Sirius was observed many times by MSX, which provides direct laboratory calibrated measurements of this star in four photometric bands from 8--22~\micron.
Thus a few bright stars have measurements directly tied to laboratory standards, but these stars are too bright for almost all of the James Webb Space Telescope \citep[JWST,][]{Gardner06} observing modes.  Hence, these measurements need to be transferred to fainter targets that are observable by JWST, and models need to be used to provide predictions at all the JWST wavelengths.

The transfer of measurements of bright stars to fainter stars requires instruments that can observe stars with a large range of flux densities with high relative accuracy.
Fortunately, this is possible both in the optical and MIR.
In the optical, the Space Telescope Imaging Spectrometer (STIS) on the Hubble Space Telescope can obtain spectra of stars with a very large range of flux densities; specifically, it can observe stars as bright as Sirius ($V = -1.46$~mag) to stars as faint as $V \sim 15$~mag \citep{Bohlin14Bright, Bohlin14}.
In the NIR and mid-infrared (MIR), this is possible using ground-based and Spitzer Space Telescope Infrared Array Camera (IRAC) observations.
For the IRAC observations, bright-star photometry can be derived from the extended wings of a star's point spread function \citep{Su21}.
This allows the relative measurement of stars with NIR and MIR flux densities from that of Sirius to stars as faint as $K_s \sim 12$~mag \citep[][G. Rieke et al., in preparation]{Rieke22}.

The design of the JWST absolute flux calibration program builds directly on the absolute calibration programs of Hubble and Spitzer which both have direct ties to the laboratory standard measurements and wavelength ranges that overlap with JWST.
The flux calibration of Hubble instruments is based on three white dwarf stars whose flux densities over the UV, optical, and NIR are set by non-LTE models fitting the Balmer lines \citep{Bohlin14}.
The absolute level of the three white dwarf spectra is set by comparison with the previously measured flux densities of Vega at 5556~\AA\ and Sirius in the MIR \citep{Bohlin20}.
For the Near Infrared Camera and Multi-Object Spectrometer (NICMOS) instrument \citep{Thompson92}, the solar analog GSPC~P330-E was also used \citep{Bohlin01, Dickinson03}.

The flux calibration of Spitzer \citep{Werner04} was independently determined for each of the three instruments.
For the Infrared Spectrograph \citep[IRS,][]{Houck04}, the calibration was based on a sample of $\sim$20 A dwarfs and $\sim$30 K giants, from which a subset of well-behaved stars were chosen as primaries \citep{Decin04, Sloan15}.
The IRS observations confirmed difficulties first seen in spectra from the Infrared Space Observatory with K giants used as primary standards in the MIR, where molecular absorption from CO, SiO, and OH were not well predicted by models \citep{Heras02, Price04, Sloan15}.
For IRAC \citep{Fazio04}, four A dwarfs and six K giants served as the primary standards.
Their flux densities were set by LTE models fit to ground-based photometry and spectroscopy \citep{Reach05, Carey12}.
For the Multiband Imaging Photometer for Spitzer \citep[MIPS,][]{Rieke04}, the flux calibration targets varied with wavelength.
At 24~\micron, the flux calibration was based on 22 A dwarfs with predicted flux densities based on MSX measurements of A stars, confirmed with solar analog stars, and then extrapolated to 24~\micron\ using a stellar atmospheric model of an A dwarf \citep{Engelbracht07, Rieke08}.
At 70~\micron, 66 stars with spectral types from B to M were used with predicted flux densities determined in the same way as at 24~\micron\ \citep{Gordon07}.
For the 53--99~\micron\ Spectral Energy Distribution mode, the calibration was based on 22 stars with K to M spectral classes from the 70~\micron\ imaging calibration sample \citep{Lu08}.
At 160~\micron\, the calibration was based on asteroids, with their flux density predictions based on the contemporaneous measurements at 24 and 70~\micron\ and a thermal emission model \citep{Stansberry07}.

The relative calibration between Hubble and Spitzer instruments and observatories are in good agreement either by the design of their calibration programs or through empirical measurements.
The calibrations of Hubble and IRAC were studied using white dwarfs, A dwarfs, and solar analogs and found to be consistent within 2\% \citep{Bohlin11}.
All the instruments on Hubble have accurate relative calibration as they are all based on the three primary white dwarfs.
The relative calibration of MIPS and IRS on Spitzer are formally tied together at 24~\micron\ \citep{Sloan15}.
\citet{Rieke08} found that the calibration between MIPS and IRAC is consistent within 1.5\% using a sample of A dwarf and solar analog stars.
A comparison of IRAC observations of the sample of A dwarfs and K giants revealed that the IRAC and IRS calibrations agree to within $\sim$1\% (Kraemer et al.\ 2022, submitted).

The goal of the JWST absolute flux calibration program is to provide flux calibration for all JWST observing modes that supports the broadest range of science.
This paper focuses on the overall program design and the calibration stars that support this goal.
The calibration stars are picked to be traceable to laboratory standards via Vega and Sirius at select optical and infrared wavelengths.
In addition, these stars can be well modeled allowing the flux densities at these select wavelengths to be transferred to all the wavelengths observed by JWST.
The details of and uncertainties in predicting the flux densities of the JWST calibration stars are beyond the scope of this paper, they can be found in \citep[][; G. Rieke et al., in preparation]{Bohlin14, Bohlin17, Bohlin20, Rieke22}.

JWST operates from the visible through the MIR (0.6--28.3~\micron), with instruments that have many capabilities similar to those on large ground-based telescopes.
This requires an approach to absolute calibration that unifies this full spectral range, and doing so with improved accuracy compared to previous work.
This approach should have significant benefits for all of optical and infrared astronomy, benefits that we will support through cross-calibrating with databases including the CALSPEC catalog\footnote{\label{calspec}https://www.stsci.edu/hst/instrumentation/reference-data-for-calibration-and-tools/astronomical-catalogs/calspec} of stellar calibrators \citep{Bohlin14} and the 2MASS infrared sky survey \citep{Skrutskie06}.

\citet{Gordon09} and \citet{Gordon12} have described early work on designing the JWST program.
This work has been refined by the JWST Absolute Flux Team, which includes experts in absolute flux calibration and the JWST instruments.
Sec.~\ref{sec:design} details the program design, Sec.~\ref{sec:sample} discusses the sample of calibration stars, and Sec.~\ref{sec:calc} describes the calculation of the calibration factors, including a discussion of conventions for photometric systems.  Finally, Sec.~\ref{sec:summary} summarizes the plan.
Future papers will give the detailed flux density predictions for each star and the observations and calibration factors for each instrument.

\section{Program Design \label{sec:design}}

The core design principle for this program is to provide a robust, cross-instrument absolute flux calibration for all JWST observations where the statistical and systematic uncertainties of the calibration can be empirically quantified.
As a result, this program will observe a network of calibration stars with all JWST instruments, both in photometric and spectroscopic modes.
The four JWST instruments are the Near-InfraRed Camera \citep[NIRCam,][]{MRieke05}, the Near-InfraRed Spectrometer \citep[NIRSpec,][]{Jakobsen22}, the Fine Guidance Sensor/Near-Infrared Imager and Slitless Spectragraph \citep[FGS/NIRISS,][]{Doyon12}, and the Mid-InfraRed Instrument \citep[MIRI,][]{Rieke15}.
The FGS/NIRISS instrument includes FGS that is used for guiding and NIRISS that is focused on science observations. 
The calibration stars have different spectral properties and a range of flux densities.
Observing a sample of calibration stars will allow for an empirical measurement of the statistical accuracy.
Observing stars with different spectral types will enable testing for systematic biases in our understanding of stars as encoded in the stellar atmosphere models used.
Stars with a range of flux densities provide empirical tests of the calibration at different flux density levels, and a robust cross-calibration between instruments having different sensitivity ranges.

Extended design principles for this program include directly connecting to the calibration of Hubble and Spitzer.
We have satisfied these principles by including in our calibration sample (see Sec.~\ref{sec:sample}) stars that were used in those flux calibration programs \citep{Reach05, Engelbracht07, Carey12, Bohlin14, Sloan15}.
In addition, over the last two decades we have worked to obtain new Hubble and Spitzer observations for the JWST calibration stars that had not already been observed \citep{Bohlin08, Bohlin10, Bohlin11, Bohlin19, Krick21}.

The required accuracies for the flux calibration of JWST are 5\% for photometry and 10--15\% for spectroscopy \citep{Gordon19}.
The budgets for these requirements include multiple terms not directly related to predicting the flux densities of calibration stars (e.g., stability of point spread functions and flat field uncertainties).
The photometry requirement is the most stringent, and there the term is a 2.8\% uncertainty in the average flux density prediction for a sample of stars \citep{Gordon19}.
The stretch goal for the absolute flux calibration is to have as accurate a calibration as possible using a reasonable amount of JWST observing time and effort.
Achieving the required level of accuracy, much less higher accuracy, requires (1) averaging the measurements of multiple calibration stars, (2) having stellar atmospheric models of high quality, (3) accurately transferring measurements tied to laboratory sources to fainter sources observable with JWST, and (4) obtaining JWST observations where measurement noise is small.
Averaging multiple stars of each type is needed as stellar atmosphere models are quite good at modeling a class of stars, but likely have a 2\% or somewhat worse accuracy modeling a specific star.
This accuracy limit is based on the results for modeling Hubble CALSPEC stars from the UV through the MIR \citep{Bohlin11, Bohlin14}.
In addition, models of specific types of stars are subject to systematic uncertainties \citep{Bohlin14}.
We estimate that we will need on the order of five calibration stars of each type to account for issues with individual stars and to quantify the systematics by type.
The number five is based on the goal of 1\% accuracy per type, allowing for one of the five stars to be found unsuitable for calibration after analysis of the JWST observations, based on experience with previous calibration projects.
In addition, we have been working with modelers to provide new stellar atmospheric models that explicitly include predictions in the NIR and MIR \citep{Bohlin17, Bohlin20}.
The models will be constrained by archival and dedicated Hubble and Spitzer observations of the JWST calibration star sample \citep{Engelbracht07, Bohlin08, Bohlin11, Bohlin14, Krick21}, providing both the transfer from laboratory measurements and allowing robust cross calibration between all three observatories.
Finally, to ensure that measurement noise is not the limiting factor, the signal-to-noise goal of the JWST observations has been set to 200 (i.e., 0.5\%).
We expect that this program will produce the level of accuracy needed to support the JWST requirements and has a strong chance of achieving the stretch calibration goal of higher accuracy.
This stretch goal would enhance the science results for many programs including supernovae, dark energy, stellar populations, and stellar structure investigations \citep{Kent09}.

\subsection{Detailed Plan Components}

\begin{deluxetable}{lrc}[tbp]
\tablewidth{0pt}
\tabletypesize{\footnotesize}
\tablecaption{JWST Observing Modes Summary\label{tab:modes}}
\tablehead{\colhead{Instrument} & \colhead{Mode} & \colhead{Details}}
\startdata
NIRCam  & Imaging & 29 filters, 0.6--5~\micron \\
        & Coronagraphy & 5 masks, 1.8--5~\micron \\
        & Slitless spectroscopy & 2.4--5~\micron\ \\
NIRSpec & MOS spectroscopy & 9 dispersers, 0.6--5.3~\micron\ \\
        & IFU spectroscopy & 9 dispersers, 0.6--5.3~\micron\ \\
        & Fixed slit spectroscopy & 9 dispersers, 0.6--5.3~\micron\ \\
NIRISS  & Wide field spectroscopy & 2 gratings, 0.8--2.2~\micron\ \\
        & Single object spectroscopy & 1 grating, 0.6--2.8~\micron \\
        & Aperture masking interferometry & 4 filters, 2.8--4.8~\micron \\
        & Imaging & 12 filters, 0.8--5.0~\micron \\
FGS     & Imaging for Guiding & open, 0.5--5.5~\micron \\
MIRI    & Imaging & 9 filters, 5--28~\micron \\
        & Coronagraphy & 4 masks, 5--12, 23~\micron \\
        & Low resolution spectroscopy & 5--12~\micron\ \\
        & IFU spectroscopy & 12 gratings, 4.9--28.8~\micron \\
\enddata
\end{deluxetable}

\begin{figure*}[tbp]
\plotone{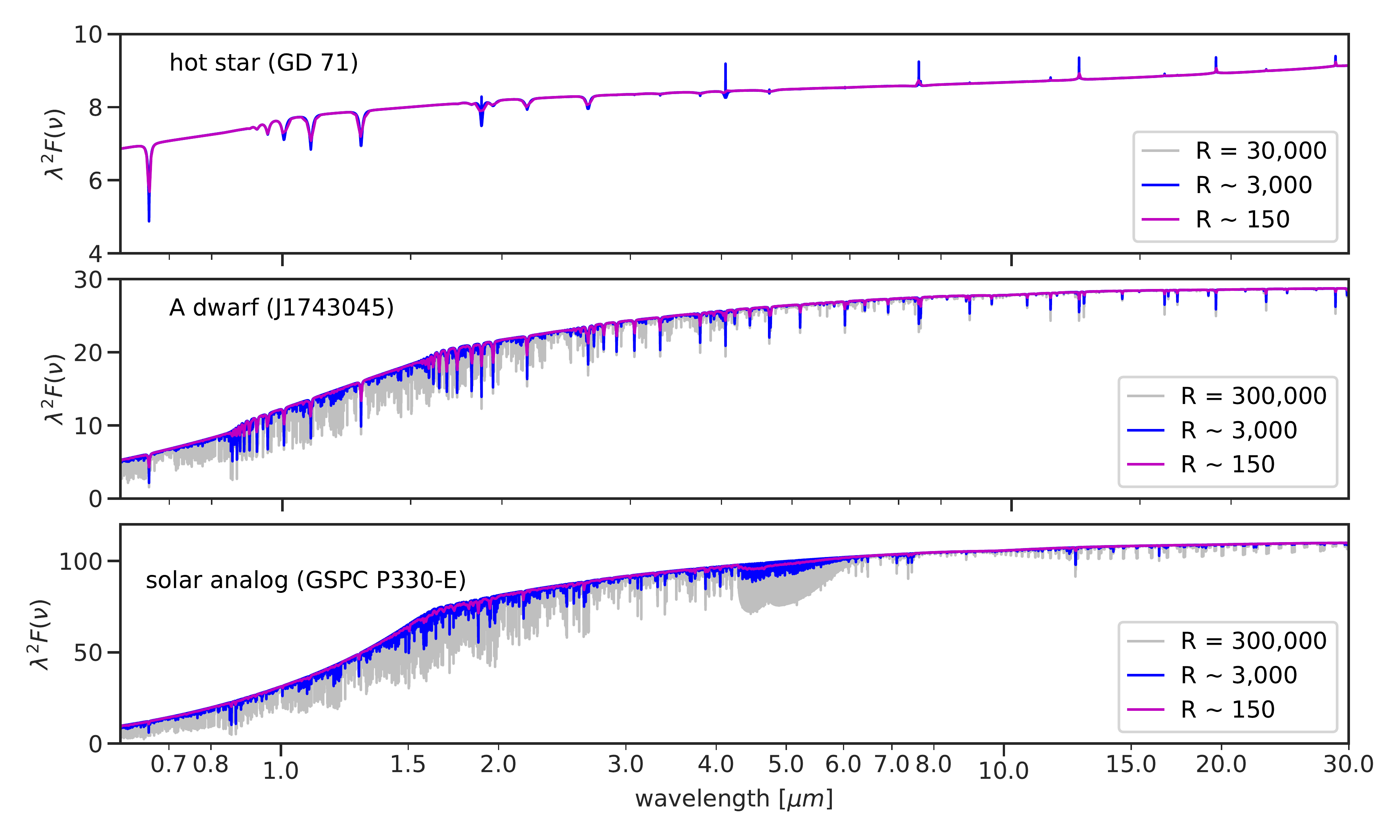}
\caption{Example spectra for each of the three types of stars included in our sample are plotted in Rayleigh-Jeans units.
The spectra are shown at spectral resolving powers of 3,000 and 150 as these match the approximate maximum and minimum spectral resolving powers of JWST spectroscopy.
For reference, the highest resolution model spectra available are shown.
\label{fig:expspec}}
\end{figure*}

JWST has four science instruments and the FGS to calibrate, and each instrument has a number of observing modes (Table~\ref{tab:modes}).
In order to efficiently calibrate all JWST observing modes and achieve the goals described above, we will observe three types of stars and have separated the observing program into five parts.
The JWST calibration stars (Sec.~\ref{sec:sample}) include hot, A dwarf, and solar analog stars, and Fig.~\ref{fig:expspec} shows example spectra of each type.
The five parts of the observing program are:  (1) observe at least one star of each type in all observing modes, (2) establish the average calibration and refine the calibration star sample by observing at least five stars of each type in a select set of modes, (3) measure the baseline instrumental uncertainty and track secular trends by observing one specific star repeatedly throughout the mission, (4) provide the transfer between subarrays and full-frame observations to give the largest dynamic range possible, and (5) investigate any dependence on brightness in the flux calibration by observing calibrator stars spanning the sensitivity range of each instrument.

{\em Part 1} focuses on the most basic goal of this program: calibration of all observing modes.
Explicitly, this means one calibration star has to be observed with every filter, optical element, and detector.
To ensure that the calibration of any one mode is robust to issues discovered after the observations, at least one star of each type is observed in each mode.
Thus, at least one of the observations should be acceptable, with the expectation that all are likely to be acceptable given the vetting carried out for the calibration star sample.

{\em Part 2} establishes the average calibration by observing a larger sample of stars with a subset of the observing modes.
The larger sample of stars contains at least five of each type and often a few more given that the sensitivities of all the observing modes do not fully overlap.
Five stars per type was picked to allow for one of the five stars to be removed from the average and still obtain a factor of two improvement in the accuracy.
This step also allows us to potentially refine the sample of stars usable for JWST absolute calibration.
Based on past experience \citep[e.g.,][]{Gordon07, Bohlin14}, a small number of the stars in the initial calibrator sample may exhibit undesirable properties (e.g., winds, weak disks, or star spots) that will only be revealed with the JWST observations themselves as these will be the most sensitive measurements of these stars ever taken at JWST wavelengths.
The subset of observing modes to be used include 10 NIRCam filters, all MIRI filters, the NIRSpec fixed-slit prism (R $\sim$ 100 from 0.6--5.3~\micron), the NIRISS SOSS grating (R $\sim$ 700 from 0.6--2.8~\micron), and the MIRI LRS prism (R $\sim$ 100 from 5--12~\micron).
This subset was chosen to provide observations over the widest range of wavelengths possible, include both photometry and spectroscopy, within a reasonable amount of observing time.

{\em Part 3} measures the minimum uncertainty of observations using repeated measurements of the same star taken over the JWST mission.
This measurement is made after correcting for all known instrumental effects.
It is an empirical repeatability measurement for a point source and is simply the scatter in the measurements in instrumental units.
This repeatability uncertainty is thought to originate in the detectors, so that the observations only need to be taken in one filter or grating per detector.
For example, such measurements made for the Spitzer IRAC 3.6 and 4.5~\micron\ and MIPS 24~\micron\ bands resulted in repeatability scatter of 0.3~\% to 0.6~\% \citep{Engelbracht07, Bohlin22}.

{\em Part 4} empirically measures any responsivity variations between full frame and subarray exposures by measuring the same star with both types of exposures.
Given the large range in sensitivity between full frame and the smallest possible subarrays, multiple stars may be used with observations that overlap for certain subarrays.
While such responsivity variations are expected to be small or non-existent, it is critical to quantify them because the large range of sensitivities across the JWST instruments requires the use of subarrays, especially for {\em Part~2} of the calibration program.

{\em Part 5} checks that the absolute flux calibration applies to a wide range of flux densities.
Correction for known instrumental non-linearities as a function of measured signal will be done as part of the standard data reduction for all the instruments. 
The goal of this part is to confirm the standard non-linearity correction and check for any possible remaining non-linearities that depend on count rate. 
Observations of calibration stars with a range of flux densities are used to measure and, if needed, correct for any brightness dependence in the flux calibration.
We will also check that the flux densities we measure for the same stars agree between different instruments with overlapping wavelengths, as disagreements can also indicate residual non-linearities.
Like {\em Part~4}, any flux density dependence to the calibration is expected to depend on detector, and hence observations are only needed for one filter or grating per detector.
This work will use many of the observations taken for {\em Parts~1} and {\em 2} with additional observations needed to fill in the areas of flux density space not already covered.

\begin{figure*}[tbp]
\epsscale{1.1}
\plotone{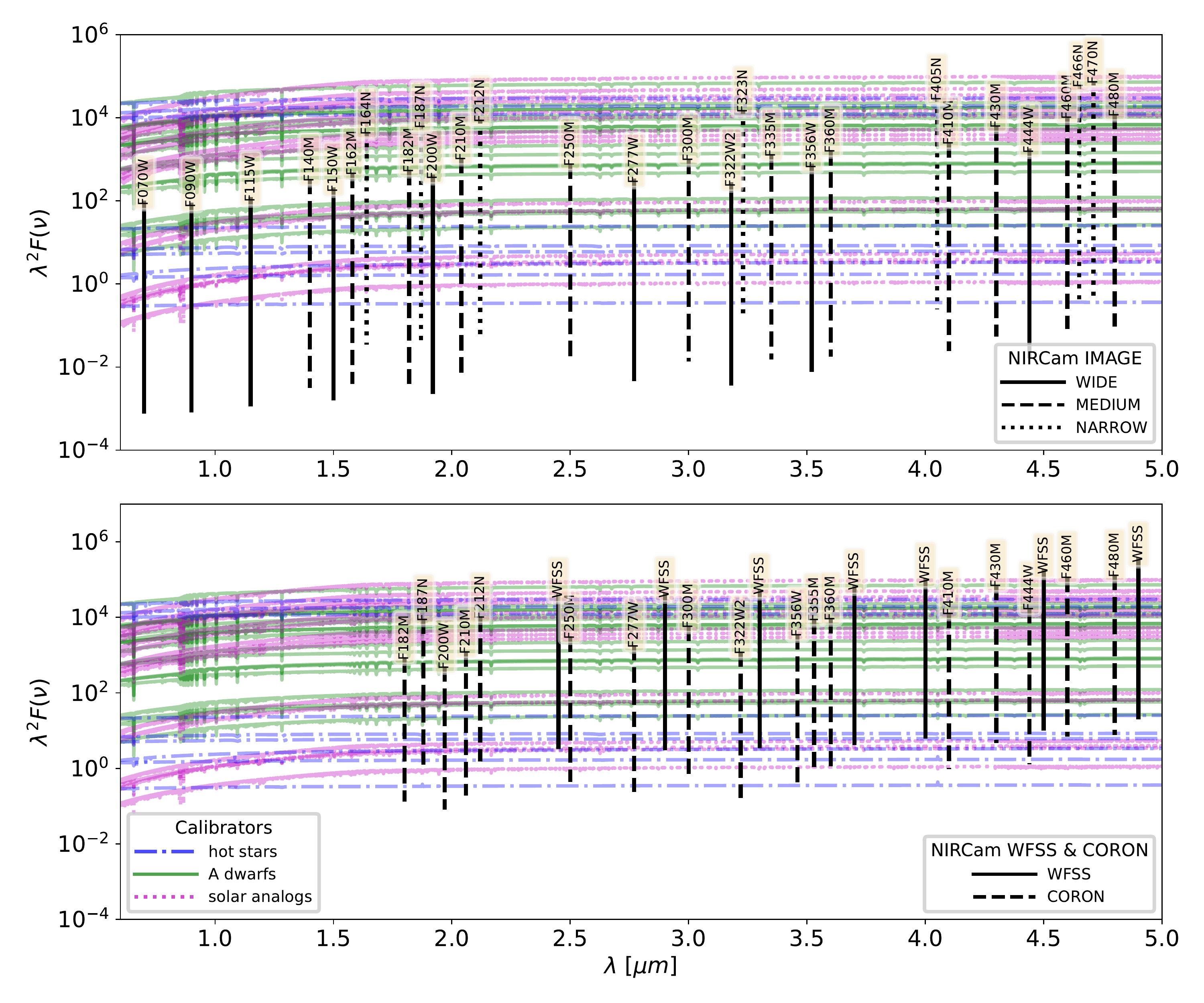}
\caption{The spectra of all the calibration stars compared to the instrumental sensitivities.  
The spectra are plotted in Rayleigh-Jeans units.
The calibration min/max ranges for the NIRCam instrument modes are plotted as vertical lines.
The wavelengths of the sensitivities of some of the modes have been shifted to minimize overlap.
\label{fig:allspec1}}
\end{figure*}

\begin{figure*}[tbp]
\epsscale{1.1}
\plotone{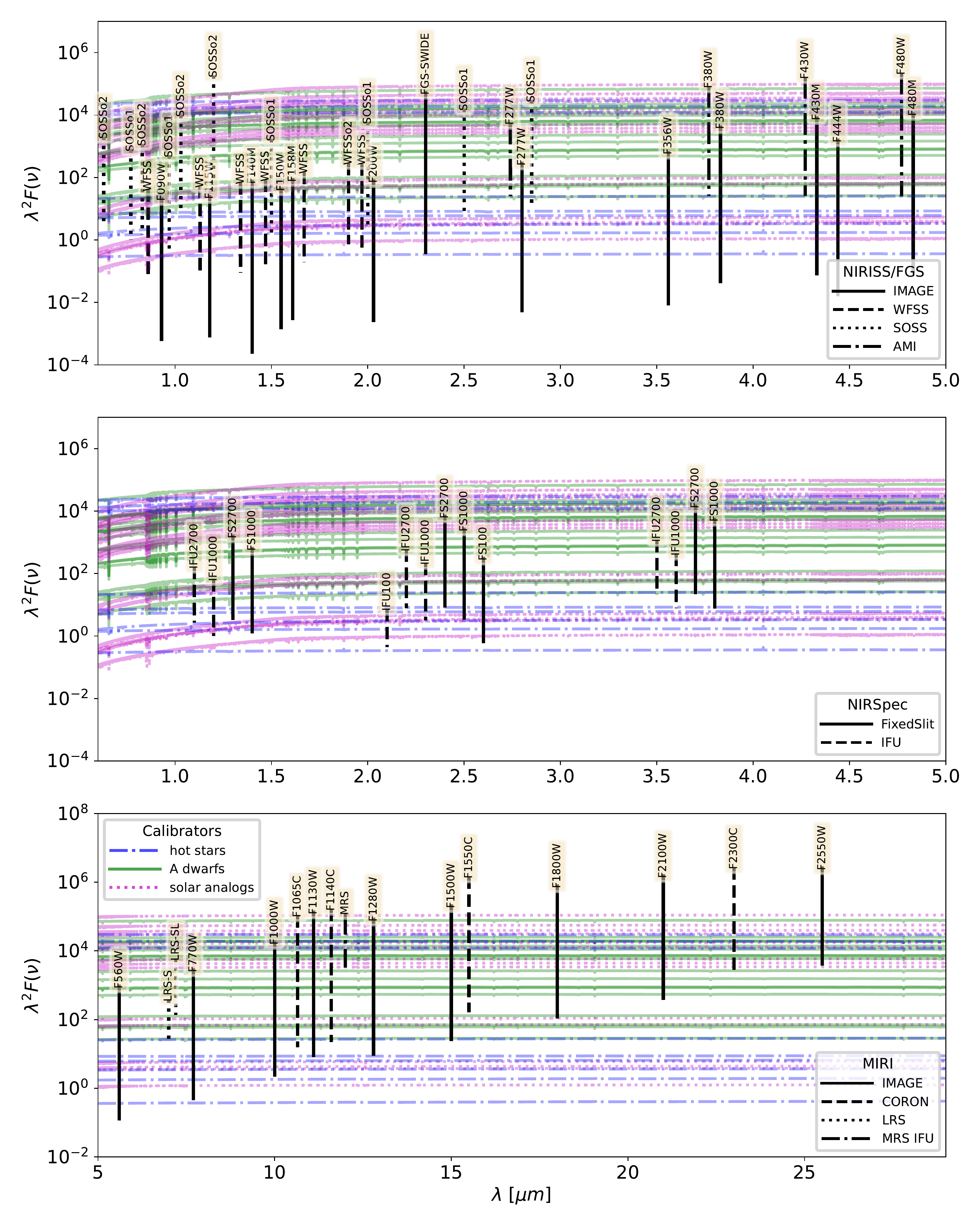}
\caption{The spectra of all the calibration stars compared to the instrumental sensitivities.  
The spectra are plotted in Rayleigh-Jeans units.
The calibration min/max ranges for the NIRISS/FGS, NIRSpec, and MIRI instrument modes are plotted as vertical lines.
The wavelengths of the sensitivities of some of the modes have been shifted to minimize overlap.
\label{fig:allspec2}}
\end{figure*}

Section~\ref{sec:sample} presents the sample of JWST calibration stars.
This sample was chosen in order to provide five stars of each type for each instrument mode.
For each instrument mode, the maximum flux density observable that does not saturate with the minimum possible integration time with the smallest supported subarray was tabulated.
Similarly, the faintest flux density observable in an exposure time of one hour with a signal-to-noise of 200 was tabulated.
Figs.~\ref{fig:allspec1} and \ref{fig:allspec2} illustrate the minimum and maximum flux densities for each mode along with the the predicted spectra of most of the calibration stars.
These figures provide a graphical summary of how the range in flux densities of our calibration stars cover the minimum and maximum flux densities targeted in this program.
Fig.~\ref{fig:allspec1} shows that our calibration stars do a good job covering the needed flux density range for NIRCam wide field slitless spectroscopy and coronagraphy, but do not cover the fainter range for many of the NIRCam imaging filters.
Fig.~\ref{fig:allspec2} shows good coverage by all NIRSpec modes, all NIRISS modes except the faint range for imaging, the one FGS mode, and all MIRI modes except for the bright range at wavelengths longer than 15~\micron.
We reach the goal of five stars of each type for most instrument modes.
However, there were some modes where this was not possible with the current sample of calibration stars.
In addition, finding both fainter and brighter calibration stars is a long-term and challenging goal.
Thus, the number of calibration stars will likely increase as more suitable stars are identified.

\subsection{Cycle 1 Plan}

The comprehensive plan detailed above will take multiple years to fully execute, both due to the need to monitor the flux calibration over the full JWST mission lifetime, and because the total amount of observation time needed is beyond what is reasonable for absolute calibration in any one year (= cycle).
One of the main driving factors in the time needed to execute this program is directly related to the extensive capabilities of the JWST instruments, as this results in many filters, gratings, etc. that require calibration.
Here we give the portions of the full plan that are planned for execution in Cycle 1, with the focus on enabling the broadest range of science.

{\em Part 1}: For the NIR instruments, one star of each of the three calibration types will be observed in every filter, grating setting, detector, etc.
For the MIR, observations of one star each from the samples of A dwarfs and solar analogs will be obtained.
Observations of the hot stars in the MIR will be obtained in future cycles once the use of such stars for MIR flux calibration is validated using observations from {\em Part~2}.

{\em Part 2}: Observations of at least three stars of each type are planned for Cycle~1.
Three stars should be enough to identify if one of the three shows non-ideal behavior.
Observations of the three types of stars will provide indications if one of the types has non-ideal behavior.
These observations will be obtained in the NIR and MIR for all three types.
In addition to the overall goals for {\em Part~2}, this will allow us to quantitatively investigate the suitability of hot stars for MIR flux calibration.

{\em Part 3}: For the majority of the observations, BD+60~1753 will be observed 10 times over the year.
For MIRI MRS observations, HD~163466 will be observed as it is significantly brighter, and will also be repeated 10 times over the year.

{\em Part 4}: One star or appropriate set of stars of different flux density levels will be observed with each detector where subarrays are supported.

{\em Part 5}: The stars picked for {\em Parts 1} and {\em 2} will be used to perform a preliminary analysis of the flux density linearity of the calibration.
In future cycles, stars with a wider range of flux densities will be observed to cover the largest range possible.

The calibration stars used for {\em Parts 1} and {\em 2} were chosen to minimize the total amount of time required in Cycle~1.
Using the same star for observations for multiple instruments reduces the overheads, especially those due to slewing to the star.
A simple algorithm was used to pick stars, with the first star being the one that was observable by most modes, then the second the next most, with this repeated until all the modes were covered.
This resulted in 3 hot stars, 8 A dwarfs, and 7 solar analogs with Table~\ref{tab:sample} indicating the specific stars.
The program IDs for the JWST programs that contain the full details of the planned observations are 1536, 1537, 1538 ({\em Parts 1, 2 \& 5}), and 1539 ({\em Part 3}).
For {\em Part 4}, the observations are spread over multiple programs including, but not limited to, 1094, 1096, and 1550.
As with all JWST data, all the observations taken for this work will be processed with the JWST pipeline and publicly available through the Mikulski Archive for Space Telescopes (MAST) archive.

\section{Calibration Star Sample \label{sec:sample}}

Our sample of calibration stars includes hot stars, A dwarfs, and solar analogs.
The use of hot stars is motivated by the fact that these stars have simple atmospheres dominated by radiative transport.
Three of the white dwarfs that are the basis of the Hubble flux calibration \citep{Bohlin14} form the core of this sample.
No similar, simple white dwarfs that are bright enough for some of JWST observation modes are available (e.g., MIRI spectroscopy), so we added bright OB dwarf stars that were used in the calibration program of the International Ultraviolet Explorer (IUE).
The A dwarfs are included because they also have simple atmospheres dominated by radiative transport, and they provided the basis of the Spitzer calibration \citep{Reach05, Engelbracht07, Sloan15}.
The final type included in the sample is solar analogs, because they are similar to our Sun, which is the star that we have the most extensive observations and knowledge of, and they are primary calibrators for ground-based NIR observations \citep{Johnson66review} and the Hubble NICMOS instrument \citep{Dickinson03}.

The sample includes multiple stars of each type.
Using an average of multiple stars of the same type has a long history in absolute flux calibration, going back to at least the work of Johnson who used a sample of A dwarf stars \citep{Johnson53, Johnson66}.
The star Vega has been used as the single calibrator for observations based on a stellar atmospheric model of an A dwarf.
Vega provides a cautionary tale, however, as it has been found to be a pole-on rapid rotator with a significant surface temperature gradient \citep{Aufdenberg06}, {\em and} it has a circumstellar debris disk with significant MIR emission above the stellar photosphere \citep{Aumann84, Su05, Bohlin14Bright}.

The criteria that make a star a good absolute flux calibrator are a mixture of properties of the star itself, the line-of-sight to the star, and the use of the star in previous absolute calibration efforts.
The stellar properties start with the star being a single star or in wide binary to avoid biases in measuring the star's flux density; close binaries require more complex models to account for complex mixed spectra and possible heating by the companion.
In addition, the star should have a spectral type that is straightforward to model.
Hot stars and A dwarfs are relatively straightforward to model as their atmospheric structures are driven by radiative transport as opposed to those dominated by convective transport.
Solar analogs can be modeled accurately as they are similar to our Sun for which we have the most detailed observations and understanding.
The star should not vary photometrically at or below the 0.25\% level ($1\sigma$), not have a circumstellar disk of gas or dust, rapid stellar rotation, or a significant wind.
Finally, the line-of-sight dust extinction should be low and well modeled.
Stars used in previous calibration efforts are always a good place to start given they have already undergone significant vetting.
A star does not have to pass every criterion, but as many as possible should be satisfied.

Table~\ref{tab:sample} gives the calibration stars in our sample.
This list may be updated as additional data are obtained including from JWST itself, but we fully expect that the majority of these sources will be suitable for JWST absolute flux calibration given the extensive vetting they have undergone.
The coordinates are from Gaia DR2 \citep{Gaia18}.
The table gives references for the spectral types, which were determined mostly using standard spectral typing techniques where optical spectra are visually compared with spectra of spectral standards.
Those stars with the "A" reference were determined from the spectral fitting of the Hubble/STIS spectroscopy to stellar atmosphere models \citep{Bohlin17}.
The $K_s$ band magnitudes are from 2MASS \citep{Skrutskie06} unless noted in the table.
The $E(B-V)$ values are from spectral fitting of the Hubble/STIS spectroscopy \citep{Bohlin17} except for NGC2506-G31 that is from \citet{Knudstrup20} and HR~7018 that is estimated from the spectral type and measured $(B-V)$ color \citep{Oja91}.
Stars that have observations planned for the first year (Cycle 1) of JWST operations are designated.
The variability is given as $1\sigma$ values for both the variability detections and upper limits.
The $1\sigma$ values were determined from the Transiting Exoplanet Survey Satellite (TESS) variability measurements by \citet{Mullally22} or from more recent equivalent TESS measurements made specifically for this paper.
The $1\sigma$ values are computed as $V_{95}/4$ or $V_{99.7}/6$, where $V_{95}$ measures the envelope of all data within $\pm 2\sigma$ of the median and $V_{99.7}$ is a similar measurement within approximately $\pm 3\sigma$.
The TESS measurements are sensitive to variations with timescales from 2 weeks to 4 min except for a few stars where the shortest timescale is 20 or 60 min \citep{Mullally22}.
For three stars (HD~2811, 18~Sco, and HD~142331) we give Hipparcos HPSCAT variability measurements for reference until new TESS variability measurements are obtained.
For 18~Sco, , the variability is from Hipparcos \citep{Adelman01}.
For the stars without existing TESS variability measurements, three have lower time resolution measurements that we will analyze in future work (HD~2811, SNAP-2, and C26202).
For all the stars, we have submitted a TESS proposal to observe them at the highest available cadence (2~min) to provide continued monitoring or their first such measurements.
For the stars not detected as variable, all had even lower upper limits that given in Table~\ref{tab:sample} for periodic variability with periods less than one day.
The notes give additional information on a star including alternative names.
All of the stars have Hubble UV/optical STIS spectroscopic observations, except for HR~7018 and NGC2506-G31.
For these two stars, a Hubble calibration proposal to obtain the STIS observations has been submitted and accepted.
Most of the sources have Spitzer IRAC and/or MIPS 24~\micron\ photometry \citep{Engelbracht07, Bohlin11, Krick21}.
The NGC2506 cluster has 19 solar analogs that are candidates for absolute flux calibration, and we have picked G31 as the best candidate as it is the most isolated (Schlawin et al., in prep.).

\begin{deluxetable*}{lccccccrcc}[tbp]
\tablewidth{0pt}
\tabletypesize{\footnotesize}
\tablecaption{Calibration Program Stars\label{tab:sample}}
\tablehead{\colhead{Name} & \colhead{RA} & \colhead{DEC} & \colhead{SpType} & \colhead{Ref} &
   \colhead{$K_s$} & \colhead{$E(B-V)$} & \colhead{Var} & \colhead{Cycle 1} & \colhead{Notes}}
\startdata
\multicolumn{9}{c}{Hot Stars} \\ \hline
$\lambda$ Lep & 05 19 34.5240 & $-$13 10 36.441 & B0.5IV & 15 & 5.09 & 0.01 & 0.09\% & & HD 34816 \\
10 Lac      & 22 39 15.6786 & +39 03 00.971 & O9V & 15 & 5.50 & 0.07 & $<$0.13\% & & HD 214680 \\
$\mu$ Col   & 05 45 59.895  & $-$32 18 23.165 & O9.5V & 15 & 5.99 & 0.01 & 0.03\% & & HD 38666 \\
G 191-B2B\tablenotemark{a} & 05 05 30.6183 & +52 49 51.921 & DA0.8 & 1 & 12.76 & 0.0 & $<$0.13\% & \checkmark & \\
GD 71\tablenotemark{a}    & 05 52 27.6197 & +15 53 13.229 & DA1.5 & 1 & 14.12 & 0.0 & $<$0.32\% & \checkmark & \\
GD 153\tablenotemark{a}   & 12 57 02.3325 & +22 01 52.634 & DA1.2 & 1 & 14.31 & 0.0 & $<$0.41\% & \checkmark & \\
LDS 749B     & 21 32 16.2328 & +00 15 14.398 & DB4 & 2 & 15.22 & 0.01 & & & TESS planned Aug 2022 \\
WD 1057+719  & 11 00 34.2433 & +71 38 02.920 & DA1.2 & 1 & 15.47  & 0.0 & $<$1.0\% & & \\
WD 1657+343  & 16 58 51.1126 & +34 18 53.321 & DA0.9 & 1 & 17.4\tablenotemark{c}  & 0.0 & $<$6.3\% & & \\ \hline
\multicolumn{9}{c}{A Dwarf Stars} \\ \hline
$\delta$ UMi\tablenotemark{b} & 17 32 12.9967 & +86 35 11.258 & A1Van  & 3 & 4.26  & 0.01 & 0.01\% & \checkmark & HD 166205 \\
HR 701\tablenotemark{b} & 02 22 54.6753 & $-$51 05 31.659 & A5V  & 4 & 5.44 & 0.01 & 0.01\% & & HD 14943 \\
$\eta^1$ Dor\tablenotemark{b} & 06 06 09.3815 & $-$66 02 22.630 & A0V  & 5 & 5.75 & 0.0 & 0.05\% & & HD 42525 \\
HR 7018\tablenotemark{b} & 18 37 33.5178 & +62 31 35.660 & A0V & 6 & 5.75 & 0.05 & 0.03\% & & HD 172728 \\
HR 5467\tablenotemark{b} & 14 38 15.2219 & +54 01 24.025 & A1V  & 6 & 5.76 & 0.0 & 0.01\% & & HD 128998 \\
HR 6514    & 17 26 04.8370 & +58 39 06.831 & A4V  & 6 & 6.15 & 0.05 & 0.10\% &  & HD 158485 \\
HD 163466\tablenotemark{b} & 17 52 25.3757 & +60 23 46.940 & A7Vm & B & 6.34 & 0.02 & 0.05\% & \checkmark & \\
HD 101452\tablenotemark{b} & 11 40 13.6509  & $-$39 08 47.674 & F0Vm & B & 6.82 & 0.02 & $<$0.02\% & & \\
HD 2811\tablenotemark{b} & 00 31 18.4899 & $-$43 36 23.000 & A3V  & 4 & 7.04 & 0.02 & 1.1\%\tablenotemark{d} & \checkmark & \\
HD 37725    & 05 41 54.3697 & +29 17 50.957 & A3V  & B & 7.90 & 0.05 & 0.13\% & & \\
HD 116405    & 13 22 45.1238 & +44 42 53.911 & A0V  & 7 & 8.48 & 0.0 & $<$0.02\% & & \\
HD 180609    & 19 12 47.1996 & +64 10 37.175 & A1V  & B & 9.12 & 0.04 & $<$0.02\% & \checkmark & \\
HD 55677    & 07 14 31.2897 & +13 51 36.786 & A2V  & 8 & 9.16 & 0.05 & 0.06\% & & \\
BD+60 1753 & 17 24 52.2772 & +60 25 50.781 & A1V  & B & 9.64  & 0.01 & $<$0.03\% & \checkmark & \\
J1757132\tablenotemark{b} & 17 57 13.2333 & +67 03 40.774 & A8Vm & B & 11.16 & 0.0 & $<$0.06\% & \checkmark & 2MASS J17571324+6703409 \\
J1802271\tablenotemark{b} & 18 02 27.1631 & +60 43 35.542 & A0V & B & 11.83 & 0.02 & $<$0.14\%& \checkmark & 2MASS J18022716+6043356 \\
J1805292\tablenotemark{b} & 18 05 29.2755 & +64 27 52.12 & A1V & B &  12.01 & 0.03 & $<$0.08\% & & 2MASS J18052927+6427520 \\
J1743045\tablenotemark{b} & 17 43 04.4857 & +66 55 01.663 & A5IIIm & B & 12.77 & 0.03 & $<$0.22\% & \checkmark & 2MASS J17430448+6655015 \\ \hline	
\multicolumn{9}{c}{Solar Analog Stars} \\ \hline
18 Sco       & 16 15 37.2704 & $-$08 22 09.982 & G2Va & 9 & 3.99  & 0.0 & 0.4\%\tablenotemark{d} & & HD 146233 \\
16 Cyg B     & 19 41 51.9732 & +50 31 03.086 & G3V & 9 & 4.66 & 0.0 & $<$0.04\% & \checkmark & HD 186427 \\
HR 6538      & 17 32 00.9923 & +34 16 16.131 & G1V & 10 & 5.05  & 0.0 & $<$0.02\% & & HD 159222 \\
HD 205905    & 21 39 10.1510 & $-$27 18 23.666 & G1.5 IV-V & 9 & 5.32  & 0.0 & 0.07\% & & \\
HD 106252    & 12 13 29.5100 & +10 02 29.889 & G0V & B & 5.93  & 0.0 & $<$0.01\% & \checkmark & \\
HD 37962     & 05 40 51.9659 & $-$31 21 03.985 & G2V & 11 & 6.27  & 0.01 & $<$0.02\% & \checkmark & \\
HD 142331    & 15 54 19.7884 & $-$08 34 49.369 & G5V & 12 & 7.13 & 0.01 & 1.8\%\tablenotemark{d} & & \\
HD 167060    & 18 17 44.1430 & $-$61 42 31.623 & G3V & 5 & 7.43 & 0.02 & $<$0.02\% & \checkmark & \\
HD 115169    & 13 15 47.3883 & $-$29 30 21.184 & G3V & 13 & 7.71 & 0.01 & $<$0.02\% & & \\
GSPC P330-E\tablenotemark{a}  & 16 31 33.8125 & +30 08 46.398 & G0V & B & 11.42 & 0.03 & $<$0.13\% & \checkmark & 2MASS J16313382+3008465 \\
GSPC P177-D  & 15 59 13.5786 & +47 36 41.905 & G0-1V & B & 11.86 & 0.04 & $<$0.17\% & \checkmark & 2MASS J15591357+4736419 \\
SNAP-2      & 16 19 46.1029 & +55 34 17.863 & G3V & B & 14.49  & 0.02 & & & 2MASS J16194609+5534178 \\
C26202      & 03 32 32.843 & $-$27 51 48.58 & F7V & A & 14.82  & 0.06 & & & 2MASS J03323287$-$2751483 \\
SF1615+001A & 16 18 14.2397 & +00 00 08.609 & G1V & A & 15.31  & 0.10 & & & 2MASS J16181422+0000086 \\
NGC2506-G31 & 08 00 14.2126 & $-$10 47 29.467 & G1V & 16 & 16.25  & 0.08 & & \checkmark & Gaia EDR3 3038045185547143936 
\enddata
\tablenotetext{a}{Hubble standard \citep{Dickinson03, Bohlin14}}
\tablenotetext{b}{Spitzer standard \citep{Reach05, Engelbracht07, Sloan15}}
\tablenotetext{c}{Calculated from a model fit to UV/optical STIS spectroscopy}
\tablenotetext{d}{Hipparcos HPSCAT variability measurement \citep{HIP97}}
\tablerefs{(1) \citet{Gianinas11}, (2) \citet{Oswalt88}, (3) \citet{Gray87}, (4) \citet{Houk78}, (5) \citet{Houk75}, (6) \citet{Cowley69}, (7) \citet{Woolley69}, (8) \citet{Fehrenbach66}, (9) \citet{Keenan89}, (10) \citet{Gray03}, (11) \citet{Gray06}, (12) \citet{Houk99}, (13) \citet{Houk82}, (14) \citet{Johnson53}, (15) \citet{Morgan55}, (16) Schlawin et al., in preparation,
(A) based on modeling by \citet{Bohlin17}, (B) MK type determined from the STIS spectra}
\end{deluxetable*}

Some stars that were initially included in our sample have been removed because they were later found to not fulfill the calibration star criteria given above.
Stars removed due to excessive variability are 2MASS~J17325264+7104431 (TESS $\sigma = 0.35$\%), 2MASS~J18083474+6927286 (TESS $\sigma = 0.41$\%), 2MASS~J18120957+6329423 (TESS $\sigma = 0.40$\%), HD~38949 (TESS $\sigma = 0.30$\%), and HD~209458 as it varies by  $\sim$2\% due to a transiting exoplanet \citep{Charbonneau00, Deming05}.
Other removals include GSPC P041-C as it has a nearby (0.57$\arcsec$) faint companion \citep{Gilliland11}, HD~27836 as it is a double star separated by 0.45$\arcsec$ in Hubble STIS observations,$^\mathrm{\ref{calspec}}$ HD~60753 as it has a STIS spectrum that is not well modeled as a single star,$^\mathrm{\ref{calspec}}$ and $\xi^2$ Cet as it is a possible spectroscopic binary and has a very late B spectral class making it too similar to A dwarf stars \citep{Johnson53, Buscombe63}.

\section{Calibration Calculation \label{sec:calc}}

The observations of the calibration stars combined with their predicted flux densities will be used to calculate the calibration factors that convert a measurement in instrumental units to physical units.
In general, the calibration factor $C_{FD}$ can be calculated for a source with an expected flux density $F$ and total integrated data numbers\footnote{Data Numbers (DN) and Analog to Digital Units (ADU) are equivalent and both are used in the JWST community.} per second (DN~s$^{-1}$) of $N$ with 
\begin{equation}
\label{eq:calfd}
C_{\mathrm{FD}} = \frac{F}{N} .
\end{equation}
Using a star as the calibration source, $N$ would have to be measured with an infinite aperture to capture all of the signal.
Hence, measurements in appropriate finite apertures are corrected to infinite apertures using aperture corrections calibrated from observations of isolated bright stars and optical models of the telescope and instrument.

The calibration factor for extended sources in surface brightness units is then
\begin{equation}
\label{eq:calsb}
C = \frac{C_{\mathrm{FD}}}{\Omega_{\mathrm{pix}}}
\end{equation}
where $\Omega_{pix}$ is the average solid angle of a pixel.
The region used to determine $\Omega_{pix}$ should be the same region of the array that was used to normalize the flat field, as the flat field corrects for the different pixel areas as well as different pixel responsivities.
These equations and those in the rest of this section are based on the work of \citet{Bohlin14}.

For JWST, the basic instrumental measurement is DN~s$^{-1}$~$\mathrm{(average\: pixel)}^{-1}$ and, thus, the natural physical unit for calibration is surface brightness.
The basic measurement is per average pixel as dividing by the normalized flat field corrects for differences in responsivity and area between pixels.
Calibrating in surface brightness units explicitly supports science for both point and extended sources.
Calibrating in surface brightness units does not compromise the accuracy of point source measurements.
Measuring a point source using the same aperture, aperture correction, and $\Omega_{pix}$ value as used for calculating the calibration factor results in an uncertainty in point source flux density that does not include the aperture correction or $\Omega_{pix}$ uncertainties.

JWST will generally calibrate observed images in mega-Janskys per steradian (MJy~sr$^{-1}$), and these are the surface brightness units common in the IR community.
For measurements integrated over spatial regions (e.g., extracted source photometry and spectrosocopy), the units will be given in Janskys (Jy).
For photometry, magnitudes in the AB and ``Vega'' systems will be provided, where the ``Vega'' system zero magnitude flux densities will be based on using Sirius as the color reference \citep{Rieke22}.

\subsection{Spectroscopy}

For spectroscopy, the calibration factors are given as a function of wavelength, where
\begin{equation}
C_{\mathrm{FD}}(\lambda) = \frac{F(\lambda)}{N(\lambda)}
\end{equation}
and
\begin{equation}
C(\lambda) = \frac{C_{\mathrm{FD}}(\lambda)}{\Omega_{\mathrm{pix}}(\lambda)} .
\end{equation}
The prediction of a source's $F(\lambda)$ includes the source spectrum and the appropriate convolution by the instrumental line-spread function.
The measurement of $N(\lambda)$ is corrected to an infinite extraction height perpendicular to the dispersion and for slit observations to an infinite slit width.
Finally, $\Omega_{\mathrm{pix}}(\lambda)$ is the average area of the pixels perpendicular to the dispersion direction at each wavelength.
In general, bandpass effects for a single wavelength pixel are small and the measured signal is attributed to the average wavelength of that pixel.
The assumption of small bandpass effects will be checked where possible.
Bandpass effects may be important, for example, in the case of observations at low spectral resolution, especially for narrow unresolved spectral lines or anywhere the sensitivity is changing rapidly across a pixel.

\subsection{Photometry}

For photometry, the bandpass effects on flux calibration can be quite significant.
The measurement in a photometric band can be calibrated as an integrated flux or as the average flux density, with the latter used more commonly.
The two common average flux density conventions are (A) to give the average flux density at the effective wavelength, and (B) to give the flux density at a fixed, reference wavelength for a reference spectral shape.
In both conventions, knowledge of the spectral shape of the source is needed to either compute the effective wavelength or correct the flux density for the difference between the actual source and reference spectral shapes.
Over the wavelength range of JWST, both conventions have been used for calibration for ground- and space-based observations \citep[e.g.,][]{Johnson65, Johnson66review, IRAS88, Bessell98, Reach05, Engelbracht07, Bohlin14}.
The following subsections give the equations for both conventions and show how it is possible to have one calibration that applies for both conventions.

\subsubsection{Convention A: Average Flux Density at $\lambda_\mathrm{eff}$}

A photometric measurement in a filter can be given as the photon-weighted average flux density, and this can be computed using
\begin{equation}
\left< F(\lambda) \right> = \frac{\int F(\lambda) R(\lambda) \lambda d\lambda}
  {\int R(\lambda) \lambda d\lambda}
\end{equation}
where $F(\lambda)$ is the flux density, $R(\lambda)$ is the bandpass function, and the integration is done over $\lambda d\lambda$ in photon units to match the detection method (i.e., \ photons $= F(\lambda)/h\nu = F(\lambda)\lambda/hc$, and $hc$ cancels out since it is in both integrals).
The bandpass function $R(\lambda)$ is the fractional transmission of the telescope and instrument including the detectors.
The effective wavelength $\lambda_\mathrm{eff}$ of this average flux density is then
\begin{equation}
\lambda_\mathrm{eff} = \frac{\int \lambda F(\lambda) R(\lambda) \lambda d\lambda}
  {\int F(\lambda) R(\lambda) \lambda d\lambda}.
\end{equation}
The number of DN detected per second is
\begin{equation}
N_\mathrm{A} = g^{-1} \int F(\lambda) R(\lambda) \lambda d\lambda
\end{equation}
where $g$ is the electronic gain in e$^-$/DN.
Thus, the calibration factor is
\begin{eqnarray}
C_{\mathrm{FD, A}} & = & \frac{\left< F(\lambda) \right>}{N_\mathrm{A}} \\
  & = & \frac{g}{\int R(\lambda) \lambda d\lambda}.  \label{eq:cuvopt}
\end{eqnarray}
Note that these quantities can also be derived using bandpass functions in photon units where the conversion to photon units is shifted from the integration to the bandpass function.

\subsubsection{Convention B: Flux Density at $\lambda_\mathrm{ref}$}

The photometry in a filter can also be given as the flux density at a reference wavelength $\lambda_\mathrm{ref}$ for a source with a reference spectral shape.
For sources that have spectral shapes different than the reference shape, then the flux density will need to be corrected for the difference in spectral shapes.
The number of DN per second for a source with $F_\mathrm{ref}(\lambda)$ is then
\begin{equation}
N_\mathrm{B} = \frac{F_\mathrm{ref}(\lambda_\mathrm{ref})}{g} \int \frac{F_\mathrm{ref}(\lambda)}{F_\mathrm{ref}(\lambda_\mathrm{ref})} R(\lambda) \lambda d\lambda
\end{equation}
and the calibration factor is
\begin{eqnarray}
C_{\mathrm{FD, B}} & = & \frac{F_\mathrm{ref}(\lambda_\mathrm{ref})}{N_\mathrm{B}} \\
  & = & \frac{g}{ \int \frac{F_\mathrm{ref}(\lambda)}{F_\mathrm{ref}(\lambda_\mathrm{ref})} R(\lambda) \lambda d\lambda }.  \label{eq:cir}
\end{eqnarray}

The flux density for an arbitrary source at the reference wavelength is
\begin{eqnarray}
F(\lambda_\mathrm{ref}) & = & \frac{C_\mathrm{FD,B} N_\mathrm{B}}{K} \\
    & = & \frac{1}{K} \frac{\int F(\lambda) R(\lambda) \lambda d\lambda}
                 {\int \frac{F_\mathrm{ref}(\lambda)}{F_\mathrm{ref}(\lambda_\mathrm{ref})} R(\lambda) \lambda d\lambda }
\end{eqnarray}
where the color correction\footnote{The term ``color correction'' is also used to describe the corrections needed to convert measurements between different photometric systems \citep[e.g.,][]{Bessell98}.} $K$ \citep{IRAS88, Reach05, Stansberry07} is
\begin{equation}
\label{eq:kcorr}
K = \frac{\int \frac{F(\lambda)}{F(\lambda_\mathrm{ref})} R(\lambda) 
                     \lambda d\lambda}
         { \int \frac{F_\mathrm{ref}(\lambda)}{F_\mathrm{ref}
                      (\lambda_\mathrm{ref})} R(\lambda) \lambda d\lambda}.
\end{equation}
Thus, the given value of a source in this convention should be divided by $K$ for the appropriate filter and spectral shape to produce the flux density at the reference wavelength of the filter.
One way to think of the color correction is that it adjusts the calibration factor to align the reference spectral shape with the current source, which results in the correct flux density at the reference wavelength.
For JWST we will provide color corrections for a reasonable range of spectral shapes different from the adopted $F_\mathrm{ref}(\lambda)$.
Custom color corrections for specific spectral shapes can be computed with the JWST bandpass functions.

The choice of $\lambda_\mathrm{ref}$ is arbitrary, but generally picked to be near the middle of the bandpass.
For JWST we adopt the pivot wavelength for $\lambda_\mathrm{ref}$ where
\begin{equation}
\lambda_{\mathrm{ref}} = \lambda_p = \left( \frac{\int R(\lambda) \lambda d\lambda}
                        {\int R(\lambda) \lambda^{-1} d\lambda} \right) ^{1/2} .
\end{equation}
The pivot wavelength has the useful properties of being source independent, and the associated pivot frequency is directly related \citep{Bohlin14}.

\subsubsection{Conversion between Conventions}

For JWST, we could pick one of the two conventions and provide conversion factors between them for every filter.
Fortunately, it is possible to pick $F_\mathrm{ref}(\lambda)$ for convention B that allows a single calibration factor per filter to work for both conventions.
This can be shown by examining the equation to convert between the two conventions.
The multiplicative conversion from convention A to convention B is the ratio of equations \ref{eq:cir} and \ref{eq:cuvopt} and is
\begin{equation}
R_\mathrm{AB} = \frac{C_\mathrm{FD, B}}{C_\mathrm{FD, A}} = 
   \frac{C_\mathrm{B}}{C_\mathrm{A}} =
   \frac{\int R(\lambda) \lambda d\lambda}
  {\int \frac{F_\mathrm{ref}(\lambda)}{F_\mathrm{ref}(\lambda_\mathrm{ref})} R(\lambda) \lambda d\lambda}.
\end{equation}
This equation is equivalent to the color correction for convention B (Eq.~\ref{eq:kcorr}) where $F(\lambda) = \mathrm{const}$.
This clue shows that convention A is mathematically equivalent to convention B with $F_\mathrm{ref}(\lambda) = \mathrm{const}$.
For JWST we adopt $F_\mathrm{ref}(\lambda) = \mathrm{const}$, resulting in the same calibration factors for both conventions A and B (i.e., $R_\mathrm{AB} = 1$).

\section{Summary \label{sec:summary}}

We have presented the motivation and details of the absolute flux calibration program for JWST.
This unified, efficient program for all JWST instruments has the goal of providing accurate absolute and relative calibration.
We designed this program to quantify and minimize the statistical and systematic uncertainties through observations of three different types of stars and multiple stars of each type having a range of flux densities.
The three types are hot stars, A dwarfs, and solar analogs, and the program stars have been extensively vetted using Hubble, Spitzer, and TESS observations along with other ancillary data.
The program includes observations to calibrate every instrument mode, improve the vetting of the program stars using the JWST observations themselves, measure the instrument repeatability for each detector, measure the calibration between subarray and full frame observations, and confirm that observations of bright and faint sources have the same calibration.
The program calibration stars have a wide range of flux densities, providing good coverage of the minimum and maximum observable range except for the faint flux densities for NIRCam and NIRISS imaging and bright flux densities for MIRI observations beyond 15~\micron.
Carrying out the full objectives of this program will require observations taken throughout the JWST mission, and we give the specifics of the planned first year of observations.
How the calibration factors will be calculated is given for photometry and spectroscopy.
For JWST we will use a photometric calibration convention that directly supports the two commonly used conventions.
Future papers on the JWST absolute flux program and related efforts will refine the detailed flux predictions for all stars and provide calibration factors for each for the many modes for each JWST instrument.

\bibliography{jabsflux}{}
\bibliographystyle{aasjournal}

\end{document}